\documentclass[pre,superscriptaddress,showpacs,twocolumn]{revtex4}

\usepackage{graphicx}
\usepackage{graphicx,amsfonts}
\usepackage{epsfig,amsmath}
\usepackage{verbatim}

\begin{document}

\newcommand{\atanh}
{\operatorname{atanh}}
\newcommand{\ArcTan}
{\operatorname{ArcTan}}
\newcommand{\ArcCoth}
{\operatorname{ArcCoth}}
\newcommand{\Erf}
{\operatorname{Erf}}
\newcommand{\Erfi}
{\operatorname{Erfi}}
\newcommand{\Ei}
{\operatorname{Ei}}

\title{Super-Aging in two-dimensional random ferromagnets} 

\author{Raja Paul}
\affiliation{BIOMS, IWR, University of Heidelberg, 69120 Heidelberg, Germany} 
\author{Gr\'egory Schehr}
\affiliation{Theoretische Physik Universit\"at des Saarlandes
66041 Saarbr\"ucken Germany}
\affiliation{Laboratoire de Physique Th\'eorique (UMR du
  CNRS 8627), Universit\'e de Paris-Sud, 91405 Orsay Cedex,
  France}
\author{Heiko Rieger}
\affiliation{Theoretische Physik Universit\"at des Saarlandes
66041 Saarbr\"ucken Germany}

\date{\today}

\begin{abstract}

We study the aging properties, in particular the two-time
autocorrelations, of the two-dimensional randomly diluted Ising
ferromagnet below the critical temperature via Monte-Carlo
simulations. We find that the autocorrelation function
displays additive aging $C(t,t_w)=C_{st}(t)+C_{ag}(t,t_w)$, where the
stationary part $C_{st}$ decays algebraically. The aging part shows
anomalous scaling $C_{ag}(t,t_w)={\cal C}(h(t)/h(t_w))$, where $h(u)$ is a
non-homogeneous function excluding a $t/t_w$ scaling.
\end{abstract}

\maketitle

The phase ordering kinetics in pure systems has attracted much
attention in the last years \cite{bray_review}. A common scenario for
instance for ferromagnets after a fast quench from above to below the
ordering temperature is a continuous domain growth governed by a single
length scale that depends algebraically on the time $t_w$ after the
quench. The existence of this length scale quite frequently determines
also the scaling properties of other dynamical non-equilibrium
quantities like the two-time auto-correlation function $C(t,t_w)$, which
describes the correlations between the spin configurations at the time
$t_w$ after the quench and a later configuration at a time $t+t_w$. It
gives rise to what is called {\it simple aging} in the context of 
glassy systems \cite{aging-review}: 
$C(t,t_w)$ depends for large times $t$ and $t_w$ only on the scaling
variable $t/t_w$. This behavior is rather well established by
analytical works in various non-random models~\cite{analytical_aging},
and it has been corroborated by a large amount of numerical
work \cite{aging-review}.

Much less analytical results are available for disordered
ferromagnets, where numerical simulations thus play an important
role. A recent numerical study of the relaxational dynamics in
two-dimensional random magnets \cite{raja_growth} found evidence for a
power law growth $L(t) \propto t^{1/z}$ of the aforementioned length
scale $L(t)$. The dynamical exponent $z$ turned out to depend both on
temperature $T$ and disorder strength and to behave as $z \propto 1/T$
at low temperatures $T$. The latter is compatible with activated
dynamics of pinned domain walls over {\it logarithmic} free energy
barriers (rather than power law~\cite{huse_henley}).
The apparent existence of a single length scale that grows algebraically was
confirmed by a recent numerical work \cite{pleimling_disferro}, where it was
furthermore claimed that the response function is well described by local
scale invariance \cite{henkel_lsi}. In spite of this, the correlation function
showed systematic deviations from a simple $t/t_w$ scaling
\cite{pleimling_disferro} (although simple aging seems to
work well in $d=3$ \cite{aging_disferro_3d}). In \cite{pleimling_disferro}, the
autocorrelation was then 
compared to the scaling form $C(t,t_w)\sim t^{-x}\tilde{c}(t/t_w)$, which
usually holds at a critical 
point with $x > 0$ \cite{calabrese-gambassi}. However, a fit of this
form to the numerical data obtained in
\cite{pleimling_disferro} (and to ours as we will report below)
yielded {\it negative} exponents $x$, which is unphysical. The aim of
this paper is to suggest an alternative picture originally applied in
the context of aging experiments in glasses \cite{struik} and spin
glasses \cite{sg-exp}. 

We study the site diluted Ising model (DIM)  
defined on $2$-dimensional square lattice with periodic boundary
conditions, and described by the Hamiltonian
\begin{eqnarray}\label{def_diluted}
H = -\sum_{\langle i j \rangle} \rho_i \rho_j s_i s_j
\label{eq_Hamil}
\end{eqnarray} where $s_i = \pm 1$ are Ising spins, the
$\rho_i$'s are, quenched, identical and independent random variables
distributed according to the probability distribution $P(\rho) =
p\delta_{\rho,1} + (1- p)\delta_{\rho,0}$. Above the percolation
threshold $p > p_c$, with $p_c \simeq 0.593 $
\cite{percolation_staufer}, the equilibrium phase diagram is characterized by
a critical line $T_c(p)$ (with $T_c(p_c) = 0$) which separates
a ferromagnetic phase at low temperature $T$ from a
paramagnetic one at high $T$. Here we focus on the
relaxational dynamics of this system (\ref{def_diluted}) following a
quench in the ferromagetic phase, $T<T_c(p)$. At
the initial time $t=0$, up
and down spins are 
randomly distributed 
on the occupied sites when it is suddenly quenched below $T_c(p)$ 
where it evolves according to
Glauber dynamics (corresponding to the heat-bath algorithm)
with random sequential update, representing a discretized 
version of model A dynamics, {\it i.e.} for non-conserved order parameter.

\begin{figure}
\centering
 \includegraphics[width=\linewidth]
 {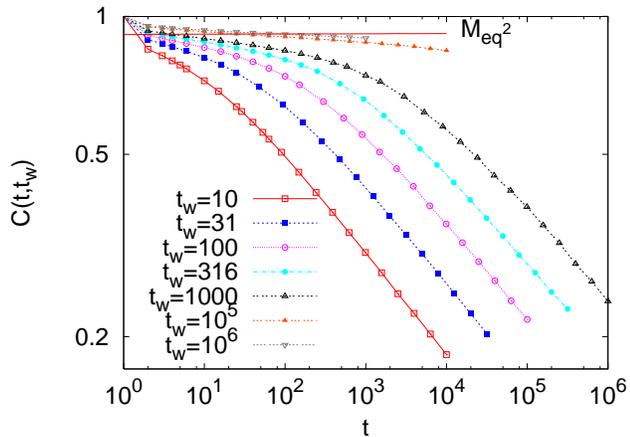}
 \caption{$C(t,t_w)$ plotted in the double logarithmic scale as a function of
 $t$ for different waiting times $t_w = 10,31,100,316,1000,10^5,10^6$. $M_{\rm
 eq}$ is the equilibrium magnetization computed with Swendsen-Wang algorothm
 (see below). Data 
 set is obtained for $p=0.75$ at temperature $T=0.7T_c$.} 
 \label{Cttw_512L_0.7Tc_0.75p_unscaled_t-tw}
\end{figure}

\begin{figure}[hb]
 \includegraphics[width=\linewidth]
 {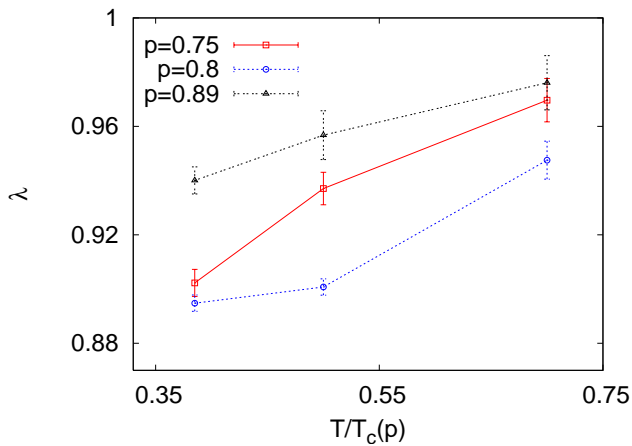}
 \caption{Exponent $\lambda$, extracted using Eq.~\ref{def_lambda}, plotted as
 function of the reduced temperature $T/T_c(p)$. These values, which violate
 the lower 
 bound $\lambda  < d/2$ has to be compared with the one for the pure
 system $\lambda_{\rm{pure}} = 1.25$.} 
 \label{T_lambda_p0.75}
\end{figure}

In the following we focus on the two-times $t>t_w$ autocorrelation
function $C(t,t_w)$ which is defined as
\begin{eqnarray}\label{def_correl}
C(t,t_w) = \frac{1}{L^2} \sum_i \overline{ \langle s_i(t+t_w) s_i(t_w) \rangle}
\end{eqnarray}
where $\langle ... \rangle$ and $\overline{...}$ stand for an average
over the thermal noise and the disorder respectively and where $L$ is the
linear system size. In our simulations, $L=512$ and $C(t,t_w)$ is 
obtained by averaging over $50$ different disorder configurations.  In
Fig.~\ref{Cttw_512L_0.7Tc_0.75p_unscaled_t-tw} we show a plot of 
 $C(t,t_w)$ as a function of the time difference $t$ and for  
different waiting times $t_w$. These data were obtained for $p=0.75$
and $T= 0.7~T_c(p=0.75)$. 
The data shown on Fig. \ref{Cttw_512L_0.7Tc_0.75p_unscaled_t-tw} on a
log-log plot suggest a power law behavior defining the
off-equilibrium exponent $\lambda$ \cite{huse_lambda}
\begin{eqnarray}\label{def_lambda}
C(t,t_w) \propto t^{-\lambda/z} \quad t \gg t_w
\end{eqnarray}
which is, as we can see on
Fig. \ref{Cttw_512L_0.7Tc_0.75p_unscaled_t-tw} weakly dependent on
$t_w$. We have checked that our simulations reproduce the well known 
values for the pure case, with $z_{\rm pure} = 2$ \cite{bray_review} and
$\lambda_{\rm pure}= 1.25$ conjectured to be exact in
Ref. \cite{fisher_spinglass}. Fig.~\ref{T_lambda_p0.75} shows a
plot of $\lambda$ as a function of $T/T_c(p)$ 
for $p =0.75, 0.8$ and $p=0.89$.  
As we can see, $\lambda(T,p)$ depends rather weakly on $T$ (in constrat with
$z$) and $p$. Besides, the obtained values   
violate the lower bound proposed in Ref. \cite{fisher_spinglass}
$\lambda \geq d/2$. Such a violation
was also obtained analytically for random field XY model in $d=2$
\cite{schehr_pre}.     

\begin{figure}
\centering
 \includegraphics[width=\linewidth]
 {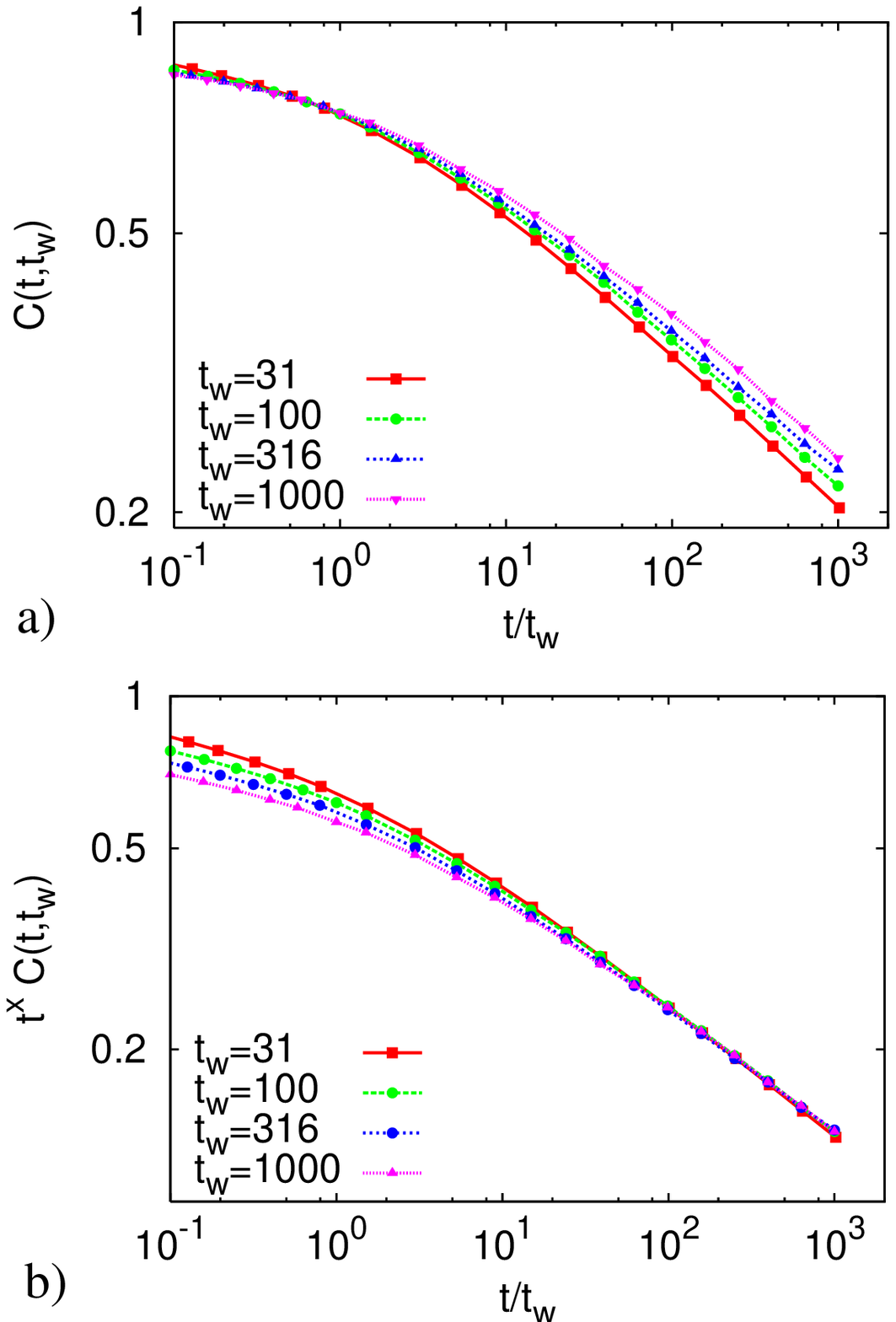}
 \caption{ {\bf a)} $C(t,t_w)$ as a function of $t/t_w$ for different $t_w =
 31,100,316,1000$. {\bf b)} $t^{x}C(t,t_w)$ as a function of
 $t/t_w$. $x=-0.04$ is 
 obtained from the 
 best data collapse. Data
 set is obtained for $p=0.75$ at temperature $T=0.7T_c$}
 \label{Cttw_512L_0.7Tc_0.75p_scaled_diff}
\end{figure}

We now focus on the scaling form of $C(t,t_w)$ as a function of both times
$t,t_w$. For non-disordered ferromagnets with purely dissipative
dynamics, one expects that $C(t,t_w)$ depends only
on the ratio $\ell(t)/\ell(t_w)$, {\it i.e.} 
\begin{eqnarray}
C_\text{pure}(t,t_w) =
 F_\text{pure}(\ell(t)/\ell(t_w)) \label{pure_scaling}
\end{eqnarray}
with $\ell(t) \propto t^{1/2}$. This has been corroborated by 
numerical simulations \cite{aging-review} as well as analytical results in 
exactly solvable limits \cite{bray_review,analytical_aging}. As shown by Paul
{\it et al.} \cite{raja_growth}, a power law domain growth is also 
observed for the present disordered system, which suggests to plot, here also,
$C(t,t_w)$ as a function of  $t/t_w$~: this is depicted in
Fig.~\ref{Cttw_512L_0.7Tc_0.75p_scaled_diff}a. 
Here one sees that this scaling form does not allow for a good
collapse of the curves for different $t_w$. The deviation from this
scaling form is indeed systematic and we have checked that the
disagreement with such a scaling persists even for larger waiting time
$t_w$. We have also obtained that simulations for other values of $p$
and $T/T_c(p)$ show the same deviations from $t/t_w$ scaling.

In Ref. \cite{pleimling_disferro} the random bond ferromagnet, which is
expected to display qualitatively the same behavior as the DIM, the
autocorrelation was compared to the scaling form
\begin{eqnarray}\label{critical_scaling}
C(t,t_w) = t^{-x(T,p)} \tilde c(t/t_w)
\end{eqnarray} 
which works well for
critical dynamics \cite{calabrese-gambassi} as well as some disordered
systems such as spin-glasses in dimension $d=3$ \cite{kisker_sg} or
an elastic line in random media
\cite{barrat_line} with a {\it positive} exponent $x(T,p) > 0$. 
However, in Ref. \cite{pleimling_disferro} a {\it negative} value
for $x(T,p)$ was obtained by fitting the data for $C(t,t_w)$ to
(\ref{critical_scaling}). We also get a good data collapse for our
data, as shown in Fig.~\ref{Cttw_512L_0.7Tc_0.75p_scaled_diff}b when
using a negative exponent $x$. The best collapse according to
Eq. (\ref{critical_scaling}) is obtained for $x=-0.04<0$ (for $p=0.75$
and $T=0.7 T_c$). The fact that $x<0$ would mean that $C(yt_w,t_w)$ grows
without bounds 
when $t_w \to \infty$ (keeping $y \gg 1$ fixed), which is
unphysical. This implies that Eq. (\ref{critical_scaling}) is not the
correct scaling form for $C(t,t_w)$, for which reason we search for an
alternative picture. First we point out that as
$t_w$ increases $C(t,t_w)$ clearly displays the formation of a
plateau (see Fig. \ref{Cttw_512L_0.7Tc_0.75p_unscaled_t-tw}). This
suggests an additive structure, as expected in the ferromagnetic phase
(and in contrast to the multiplicative scaling found at $T_c(p)$ in random
ferromagnets \cite{schehr_lux}):  
\begin{eqnarray}
C(t,t_w) = C_{st}(t) + C_{ag}(t,t_w) \label{additive}
\end{eqnarray}
such that $\lim_{t \to \infty}
C_{st}(t)=0$ and $\lim_{t \to \infty} \lim_{t_w \to \infty} C(t,t_w) =
M_{eq}^2$ where $M_{eq}$ is the equilibrium magnetization.

\begin{figure}
\centering
 \includegraphics[width=\linewidth,angle=-0]
 {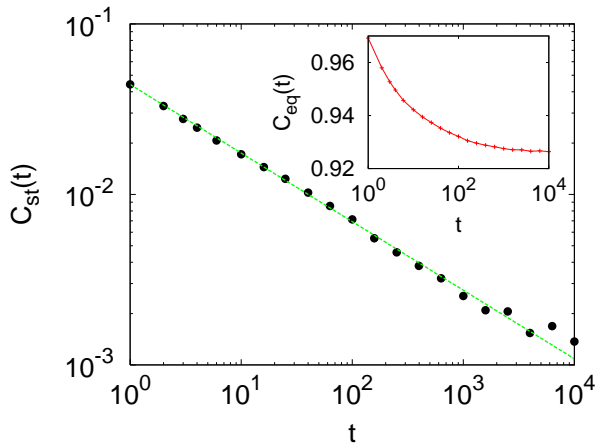}
 \caption{$C_{st}(t)$ plotted in the double logarithmic scale as a function
 of $t$. The line corresponds to $\eta = 0.4$ in
 Eq. (\ref{C_st}). {\bf Inset :} $C_{\rm eq}(t)$ as function of $t$. Data
 set is obtained for $p=0.75$ and $T=0.7\cdot T_c$.} 
 \label{Fig_eq}
\end{figure}
We first focus on the stationary component $C_{\rm st}(t)$ in
Eq. (\ref{additive}), for which analytical predictions exist relying
on droplet models \cite{huse_droplet_eq}. To study this part
numerically, we first equilibrate the system using Swendsen-Wang
algorithm \cite{swendsen_wang} and then let the system evolve
according to Glauber dynamics starting with such an equilibrated
initial configuration. We denote $C_{\rm eq}(t,t_w)$ the (equilibrium)
autocorrelation function (\ref{def_correl}) computed using this
protocol and we have checked that it is indeed
independent of $t_w$,  $C_{\rm eq}(t,t_w) \equiv C_{\rm eq}(t)$. 
\begin{figure}
\centering
 \includegraphics[width=\linewidth,angle=0]
 {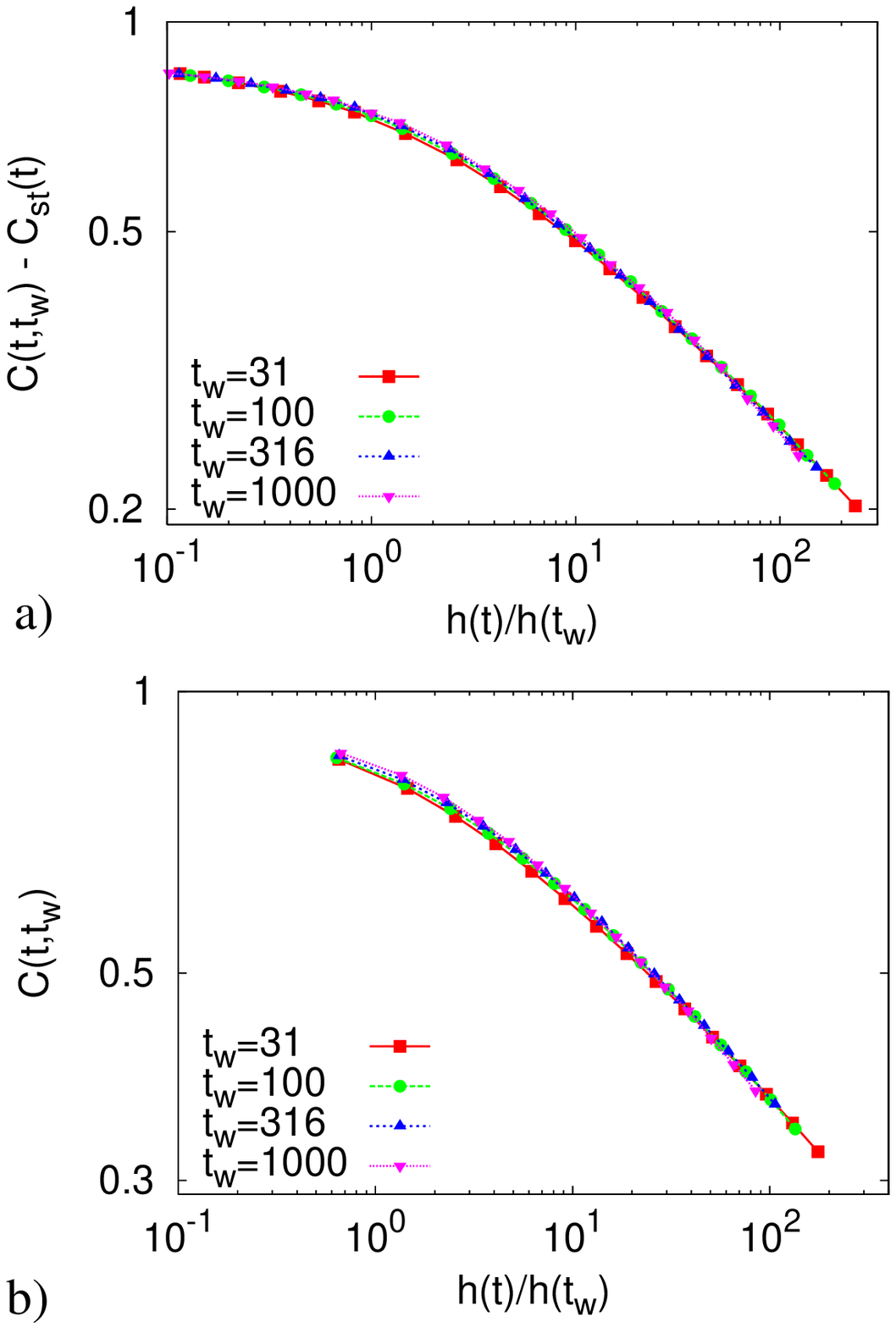}
 \caption{{\bf a)} $C_{\rm ag}(t,t_w)$ plotted as a function
 of $h(t)/h(t_w)$ with $h(t)=\exp{[t^{1-\mu}/(1-\mu)]}$
 for different $t_w = 31, 100, 316, 1000$, with $\mu = 1.035$. Data
 set is obtained for $p=0.75$ and $T=0.7\cdot T_c$. {\bf b)}
 $C(t,t_w)$ plotted as a function of $h(t)/h(t_w)$ with $\mu =
 1.042$. Data set is obtained for $p=0.75$ and $T=0.5 \cdot T_c$}   
 \label{super_aging}
\end{figure}
In the inset of Fig. (\ref{Fig_eq}) we
plot $C_{\rm eq}(t)$ as a function of $t$ for $p=0.75$ and
$T=0.7\cdot T_c$. In agreement with previous analytical predictions
\cite{huse_droplet_eq}, these data can be nicely
fitted to $C_{\rm eq}(t) = C_{\rm st}(t) + M^2_{\rm eq}$ with
$M^2_{\rm eq} = 0.925$ and a power law behavior
\begin{eqnarray}
C_{st}(t) \propto A t^{-\eta(T,p)} \label{C_st}
\end{eqnarray}
This is depicted in Fig. (\ref{Fig_eq}), where we
show a plot of $C_{st}(t)$ as a function of $t$ for $p=0.75$ and 
$T=0.7 T_c$, for these parameters, one finds $\eta = 0.40(2)$. 

We now come to the aging part $C_{ag}(t,t_w)$ in (\ref{additive}), by
first noticing that simple aging also does not hold for
$C_{ag}(t,t_w)$. Inspired by a picture originally suggested in the
context of aging experiments in glasses \cite{struik} and spin glasses
\cite{sg-exp}, and also occurring within the analytical solution
of the non-equilibrium dynamics mean field spin glasses
\cite{aging-review}, we use a form that generalizes 
Eq. (\ref{pure_scaling}):
\begin{eqnarray}
C_{ag}(t,t_w) \simeq {\cal C}(h(t)/h(t_w)) \label{mf_scaling}
\end{eqnarray}
A widely used form for $h(u)$, which we choose here, is $h(u) =
\exp{[u^{1-\mu}/(1-\mu)]}$ where 
$\mu$ allows to interpolate between super-aging ($\mu > 1$) and
sub-aging ($\mu < 1$) via simple aging ($\mu =1$). In
Fig.~\ref{super_aging}a, we show that this form with $\mu = 1.035$
allows for a nice collapse of the curves presented in
Fig.~\ref{Cttw_512L_0.7Tc_0.75p_unscaled_t-tw} for different $t_w$,
corresponding to $p=0.75$ and $T=0.7T_c$. We point out that a good
data collapse is also obtained (with the same exponent $\mu$) when
$C_{st}(t)$ is not substracted. In
Fig.~\ref{super_aging}b we show a plot of
$C(t,t_w)$ as a function of $h(t)/h(t_w)$ for $p=0.75$ and
$T=0.5T_c$. For this temperature, the best data collapse is obtained
for a larger value of $\mu = 1.042$, which suggests that $\mu$ is a
decreasing function of $T$ (and one expects $\mu \to 1$ for $T \to
T_c$).

In Ref. \cite{berthier_aging}, such a super-aging behavior -- with
comparable values of $\mu$ -- was also observed in the $4d$ Edwards
Anderson spin glass. There it was argued that super-aging is
consistent with a growth law $t(L)$ of the form
\begin{eqnarray}
t(L) \simeq \tau_0 L^{z_c}
\exp{\left({\Upsilon(T)L^\psi}/{T} \right)} \label{activated_growth}
\end{eqnarray}
where $z_c$ is the dynamic critical exponent (and here $z_c =
2.1667(5)$ \cite{nightingale_z}), $\psi$ the barrier exponent and
$\Upsilon(T)$ a typical free energy scale (vanishing at $T_c$). If one
assumes $h(t) = L(t)$ in Eq. (\ref{mf_scaling}) with $t(L)$ as in
Eq. (\ref{activated_growth}) then one can identify $\psi/z_c =
(\mu-1)$ \cite{berthier_aging}. In our case this would give a
$T$-dependent barrier exponent $\psi$ (see
Fig. \ref{Cttw_512L_0.7Tc_0.75p_unscaled_t-tw}). In 
addition, the values obtained for $\psi$ from that relation 
are different from the exact value $\psi = 1/4$ \cite{huse_henley}.

To conclude, we have performed a detailed numerical study of the
autocorrelation function during the coarsening dynamics of diluted
Ising ferromagnets in dimension $d=2$. Our data show clear deviations
from a simple $t/t_w$ scaling, which were also observed in a recent
work on a random ferromagnet in $d=2$
\cite{pleimling_disferro}. However, attempts to fit the data to the
simple scaling form as in Eq. \ref{critical_scaling} leads, as in
Ref. \cite{pleimling_disferro}, to $x < 0$, which is unphysical.  Here
we proposed an alternative way of describing the dynamics in terms of
super-aging: this allows for a consistent description of the
autocorrelation function in disordered ferromagnets. If our 
results reflect correctly the asymptotic scaling behavior of the
autocorrelation in 2d disordered ferromagnets 
one would thus conclude that it is not accurately described by local scale
invariance as in Ref. \cite{lsi_correl}.  With regards to a
recent experimental study of super-aging in spin glasses
\cite{zotev_sg}, it would be interesting to understand whether the
super-aging behavior we find is related to the choice of the initial
conditions for the dynamics.

\acknowledgments
GS acknowledges the financial support provided
through the European Community's Human Potential Program under
contract HPRN-CT-2002-00307, DYGLAGEMEM. GS and RP thank A. Coniglio,
M. Henkel and M. Pleimling for useful 
correspondence and discussions.

\end{document}